\documentstyle[graphicx,12pt]{article}
\begin{document}

\small
\hoffset=-1truecm
\voffset=-2truecm
\title{\bf The Casimir effect for a cavity in the spacetime with an extra dimension}
\author{Cheng Hong-Bo\footnote {E-mail address:
hbcheng@public4.sta.net.cn}\\
Department of Physics, East China University of Science and
Technology,\\ Shanghai 200237, China}

\date{}
\maketitle

\begin{abstract}
We reexamine the Casimir effect for the rectangular cavity with
two or three equal edges in the presence of compactified universal
extra dimension. We derive the expressions for the Casimir energy
and discuss the nature of Casimir force. We show analytically the
extra-dimension corrections to the standard Casimir effect to put
forward a new method of exploring the existence of extra
dimensions of the Universe.
\end{abstract}
\vspace{8cm} \hspace{1cm}
PACS number(s): 11.10.Kk, 04.62.+v

\newpage

Kaluza and Klein presented the thought that our spacetime has more
than three spatial dimensions over decades ago in order to unify
gravity and classical electrodynamics [1, 2]. Now the idea of
extra dimensions attracts more and more interests of the physical
community. It is fundamental for string theory to invoke seven
additional spatial dimensions to unify the quantum mechanics and
gravity. Whether the order of the compactification scale of the
extra dimensions is much larger or smaller remains unexplained.
Some models of string theory predicted that the typical size of
universal extra dimensions (UXDs) is about $10^{-35}m$ which is
beyond our recent experimental reach [3, 4]. Recently the
Kaluza-Klein idea is considered in the context of the brane
paradigm, which could lead to a solution to the hierarchy problem
[5-7]. The brane world scenario [8, 9] stipulates that the
non-gravitational physics like Standard Model is confined to a
four-dimensional Universe called brane that is embedded in a
$(4+n)$-dimensional spacetime with $n$ compactified extra
dimensions. Only gravitational field is allowed to propagate
through the whole spacetime while the possibility that the
additional dimensions may be much larger than we thought in the
past was showed.

The Casimir effect is essentially a direct consequence of quantum
field theory because of a change in the spectrum of vacuum
oscillations when the quantization volume is bounded, or some
background field is presented. The effect becomes one of the most
interesting macroscopic manifestations of the nontrivial
properties of the physical vacuum. The Casimir effect has been a
subject of extensive research [10-16]. Nowadays the Casimir effect
is used as a powerful method for the research on new physics
beyond the standard model. We apply the tool to a lot of topics
such as the right order of the cosmological constant, extra
dimensions, etc..

It is necessary to explore the dimensionality and structure of our
Universe in various directions. If our world has six or seven
additional spatial dimensions, the new phenomena related to the
existence of these dimensions should be observed. A lot of
progresses of studying the UXDs by means of Casimir effect were
made [17, 18]. The authors of [18] compared the expression of
Casimir force for parallel plates in the presence of one extra
dimension with the experimental results and the approximate values
of the compactification scale of extra dimension were obtained.

In this paper, we investigate the extra-dimension correction to
the Casimir effect for a cavity in detail to put forward a new
method to explore the existence of extra dimensions of our
Universe. There will be influence from UXD on the nature of
Casimir force between the configuration boundaries that confine
the field in the spacetime with extra dimensions. It was shown
that the sign of Casimir energy for a rectangular cavity with two
or three equal edges in four-dimensional Minkowski spacetime is
positive or negative respectively for the case of massless scalar
field [14, 19]. Here we also discuss the Casimir effect of this
kind of field satisfying Dirichlet boundary conditions within a
rectangular with two or three equal edges in the spacetime with
one extra dimension for the sake of comparing results of [14, 19]
and recent ours. Having derived the total energy, we regularize
the energy by means of Epstein zeta function to obtain the Casimir
energy. We find that the Casimir force is always repulsive if the
lengths of edges are sufficiently large with respect to the size
of UXD.

In order to derive and calculate the Casimir energy, in the
Kaluza-Klein theory we chose the Lagrangian density of a simple
model of scalar field in $(4+n)$-dimensional spacetime like,

\begin{equation}
{\cal L}=\frac{1}{2}\partial_{A}\Phi\partial^{A}\Phi
\end{equation}

\noindent with the metric $(+----)$. Here the field
$\Phi(x^{A})=\Phi(x^{\mu},y)$, $A=0,1,2,3,5$ is the function of
four-dimensional coordinates $x^{\mu}$ with $\mu=0,1,2,3$ as well
as an extra coordinate $y$. In the five-dimensional spacetime like
$M^{4}\times S^{1}$, with the compactification of extra dimension
on a circle $S^{1}$ of radius $L$, the field $\Phi(x^{A})$ can be
expanded in the harmonics as follow,

\begin{equation}
\Phi(x^{\mu},y)=\sum_{n=-\infty}^{\infty}\phi_{n}(x^{\mu})e^{\frac{iny}{L}}
\end{equation}

\noindent By substituting the expansion (2) into (1), the
Lagrangian density (1) becomes,

\begin{equation}
{\cal
L}=\frac{1}{2}\partial_{\mu}\phi_{0}\partial^{\mu}\phi_{0}
+\sum_{k=1}^{\infty}(\partial_{\mu}\phi_{k}\partial^{\mu}\phi_{k}^{*}
+\frac{k^{2}}{L^{2}}\phi_{k}\phi_{k}^{*})
\end{equation}

\noindent where $L$ is the radius of UXD as mentioned above. The
scalar fields $\phi_{k}$ satisfy the free Klein-Gordon equations,

\begin{equation}
(\partial_{\mu}\partial^{\mu}-\frac{k^{2}}{L^{2}})\phi_{k}(x)=0
\end{equation}

\noindent here $k=0,1,2,\cdot\cdot\cdot$. The fields are confined
in the interior of rectangular cavity $\Omega$ with $p$ edges. We
consider the case of Dirichlet boundary conditions, i.e.,
$\phi_{k}(x)|_{\partial\Omega}=0$. The total energy density of the
fields in the interior of $\Omega$ is thus given by,

\begin{equation}
\varepsilon_{p}=\int
d^{3-p}k\sum_{\{n\}=1}^{\infty}\sum_{n=0}^{\infty}\frac{1}{2}\sqrt{k_{T}^{2}
+(\frac{n_{1}\pi}{R_{1}})^{2}+(\frac{n_{2}\pi}{R_{2}})^{2}
+\cdot\cdot\cdot+(\frac{n_{p}\pi}{R_{p}})^{2}+\frac{n^{2}}{L^{2}}}
\end{equation}

\noindent Following [11, 14], Eq. (5) becomes,

\begin{eqnarray}
\varepsilon_{p}=-\frac{1}{2^{4-p}}\Gamma(-\frac{4-p}{2})
E_{p+1}(\frac{\pi^{2}}{R_{1}^{2}},\frac{\pi^{2}}{R_{2}^{2}},
\cdot\cdot\cdot,\frac{\pi^{2}}{R_{p}^{2}},\frac{1}{L^{2}};-\frac{4-p}{2})\nonumber\\
-\frac{1}{2}\Gamma(-\frac{4-p}{2})E_{p}(\frac{\pi^{2}}{R_{1}^{2}},\frac{\pi^{2}}{R_{2}^{2}},
\cdot\cdot\cdot,\frac{\pi^{2}}{R_{p}^{2}};-\frac{4-p}{2})
\end{eqnarray}

\noindent where Epstein zeta function
$E_{p}(a_{1},a_{2},\cdot\cdot\cdot,a_{p};s)$ is defined as,

\begin{equation}
E_{p}(a_{1},a_{2},\cdot\cdot\cdot,a_{p};s)=
\sum_{\{n\}=1}^{\infty}(\sum_{j=1}^{p}a_{j}n_{j}^{2})^{-s}
\end{equation}

\noindent $\{n\}$ stands for a short notation of
$n_{1},n_{2},\cdot\cdot\cdot,n_{p}$, $n_{a}$ a positive integer.

In addition, the Casimir force can be denoted as,

\begin{equation}
f_{C}=-\frac{\partial\varepsilon_{p}}{\partial L_{p}}
\end{equation}

We consider $p=2$ case in which the fields are confined in the
interior of cavity with two equal edges such as $R_{1}=R_{2}=R$ in
the presence of one compactified universal extra dimension. For
the case of four-dimensional Minkowski spacetime, the Casimir
energy of two-edge hypercube keeps positive [14, 19]. In the
four-dimensional spacetime with one extra dimension, the
expression (6) is reduced to,

\begin{equation}
\varepsilon_{2}=-\frac{1}{4}\Gamma(-1)E_{3}(\frac{\pi^{2}}{R^{2}},\frac{\pi^{2}}{R^{2}},\frac{1}{L^{2}};-1)
-\frac{1}{4}\Gamma(-1)E_{2}(\frac{\pi^{2}}{R^{2}},\frac{\pi^{2}}{R^{2}};-1)
\end{equation}

\noindent By regularizing Eq.(9), we obtain the Casimir energy,

\begin{eqnarray}
\varepsilon_{2}=-\frac{1}{8\pi^{\frac{5}{2}}}\Gamma(\frac{3}{2})\zeta(3)\frac{1}{L^{2}}
+\frac{1}{4\pi^{4}}\mu\Gamma(2)\zeta(4)\frac{1}{L^{2}}
+\frac{1}{2\pi^{\frac{1}{2}}\mu^{\frac{1}{2}}}\frac{1}{L^{2}}
\sum_{n_{2},n=1}^{\infty}(\frac{n}{n_{2}})^{\frac{3}{2}}K_{\frac{3}{2}}(2\mu
n_{2}n)\nonumber\\
-\frac{1}{8\pi^{\frac{11}{2}}}\mu^{2}\Gamma(\frac{5}{2})\zeta(5)\frac{1}{L^{2}}
-\frac{1}{2\pi}\frac{1}{L^{2}}\sum_{n_{2},n=1}^{\infty}(\frac{n}{n_{2}})^{2}K_{2}(2\mu
n_{2}n)\hspace{3.5cm}\nonumber\\
-\frac{1}{2\mu^{2}}\frac{1}{L^{2}}\sum_{k=0}^{\infty}\frac{16^{-k}}{k!}\prod_{j=1}^{k}[9-(2j-1)^{2}]
\hspace{6.3cm}\nonumber\\
\times\sum_{n_{1},n_{2},n=1}^{\infty}n_{1}^{-k-2}(\pi^{2}n_{2}^{2}+\mu^{2}n^{2})^{-\frac{k-1}{2}}
\exp[-2n_{1}(\pi^{2}n_{2}^{2}+\mu^{2}n^{2})^{\frac{1}{2}}]\hspace{1cm}\nonumber\\
+\frac{1}{2\pi^{\frac{1}{2}}\mu^{2}}\Gamma(\frac{3}{2})\zeta(3)\frac{1}{L^{2}}
-\frac{1}{2\pi\mu^{2}}\frac{1}{L^{2}}\Gamma(2)\zeta(2)\beta(2)\hspace{4.7cm}
\end{eqnarray}

\noindent where

\begin{equation}
\mu=\frac{R}{L}
\end{equation}

\noindent and

\begin{equation}
\beta(s)=\sum_{n=0}^{\infty}\frac{(-1)^{n}}{(2n+1)^{s}}
\end{equation}

\noindent If the lengths of cavity are large enough like $\mu>>1$,
the Casimir energy density (10) becomes,

\begin{equation}
\varepsilon_{2}(\mu>>1)=-\frac{1}{8\pi^{\frac{11}{2}}}\mu^{2}\Gamma(\frac{5}{2})\zeta(5)\frac{1}{L^{2}}
\end{equation}

\noindent In the case of $\mu<<1$, then the energy density can be
denoted as,

\begin{eqnarray}
\varepsilon_{2}(\mu<<1)=
-\frac{1}{2\mu^{2}}\frac{1}{L^{2}}\sum_{k=0}^{\infty}\frac{16^{-k}}{k!}\prod_{j=1}^{k}[9-(2j-1)^{2}]
\hspace{6.3cm}\nonumber\\
\times\sum_{n_{1},n_{2},n=1}^{\infty}n_{1}^{-k-2}(\pi^{2}n_{2}^{2}+\mu^{2}n^{2})^{-\frac{k-1}{2}}
\exp[-2n_{1}(\pi^{2}n_{2}^{2}+\mu^{2}n^{2})^{\frac{1}{2}}]\hspace{1cm}\nonumber\\
+\frac{1}{2\pi^{\frac{1}{2}}\mu^{2}}\Gamma(\frac{3}{2})\zeta(3)\frac{1}{L^{2}}
-\frac{1}{2\pi\mu^{2}}\frac{1}{L^{2}}\Gamma(2)\zeta(2)\beta(2)\hspace{4.7cm}\nonumber\\
>0\hspace{12cm}
\end{eqnarray}

\noindent We show the curves of the Casimir energy density for a
rectangular cavity with two equal edges in the spacetime with one
UXD in Figure 1. Furthermore, numerical calculation show that
there exists a particular critical ratio $\mu_{C}=5.526$ such that
the energy density $\varepsilon_{2}<0$ if $\frac{R}{L}>\mu_{C}$
while in the opposite case $\frac{R}{L}<\mu_{C}$, the energy
density remains positive. At $\frac{R}{L}>\mu_{C}$, the nature of
Casimir force is repulsive.

In this section we consider the Casimir effect for a cube with
equal edges like $R_{1}=R_{2}=R_{3}=R$ in the spacetime with one
UXD. The total energy density for the system can be derived as,

\begin{equation}
\varepsilon_{3}=\frac{1}{2}
E_{4}(\frac{\pi^{2}}{R^{2}},\frac{\pi^{2}}{R^{2}},\frac{\pi^{2}}{R^{2}},\frac{1}{L^{2}};-\frac{1}{2})
+\frac{1}{2}E_{3}(\frac{\pi^{2}}{R^{2}},\frac{\pi^{2}}{R^{2}},\frac{\pi^{2}}{R^{2}};-\frac{1}{2})
\end{equation}

\noindent By regularizing the expression (15), we obtain the
Casimir energy density,

\begin{eqnarray}
\varepsilon_{3}=\frac{1}{182}\frac{1}{L}-\frac{3}{32\pi^{\frac{7}{2}}}\mu\Gamma(\frac{3}{2})\zeta(3)\frac{1}{L}
-\frac{1}{8\pi}\frac{1}{L}\sum_{n_{3},n=1}^{\infty}\frac{n}{n_{3}}K_{1}(2\mu
n_{3}n)\hspace{2.1cm}\nonumber\\
+\frac{3}{32\pi^{5}}\mu^{2}\Gamma(2)\zeta(4)\frac{1}{L}
+\frac{\mu^{\frac{1}{2}}}{4\pi^{\frac{3}{2}}}\frac{1}{L}
\sum_{n_{3},n=1}^{\infty}(\frac{n}{n_{3}})^{\frac{3}{2}}K_{\frac{3}{2}}(2\mu
n_{3}n)\hspace{2.5cm}\nonumber\\
+\frac{1}{8\pi^{\frac{1}{2}}}\frac{1}{\mu}\frac{1}{L}\sum_{k=0}^{\infty}\frac{16^{-k}}{k!}
\prod_{j=1}^{k}[4-(2j-1)^{2}]\hspace{5.1cm}\nonumber\\
\times\sum_{n_{2},n_{3},n=1}^{\infty}n_{2}^{-k-\frac{3}{2}}(\pi^{2}n_{3}^{2}+\mu^{2}n^{2})^{-\frac{2k-1}{4}}
\exp[-2n_{2}(\pi^{2}n_{3}^{2}+\mu^{2}n^{2})^{\frac{1}{2}}]\nonumber\\
-\frac{1}{32\pi^{\frac{13}{2}}}\mu^{3}\Gamma(\frac{5}{2})\zeta(5)\frac{1}{L}
-\frac{\mu}{8\pi^{2}}\frac{1}{L}\sum_{n_{3},n=1}^{\infty}(\frac{n}{n_{3}})^{2}K_{2}(2\mu
n_{3}n)\hspace{2.3cm}\nonumber\\
-\frac{1}{8\pi\mu}\frac{1}{L}\sum_{k=0}^{\infty}\frac{16^{-k}}{k!}\prod_{j=1}^{k}[9-(2j-1)^{2}]\hspace{5.3cm}\nonumber\\
\times\sum_{n_{2},n_{3},n=1}^{\infty}n_{2}^{-k-2}(\pi^{2}n_{3}^{2}+\mu^{2}n^{2})^{-\frac{k-1}{2}}
\exp[-2n_{2}(\pi^{2}n_{3}^{2}+\mu^{2}n^{2})^{\frac{1}{2}}]\nonumber\\
-\frac{1}{4\pi^{\frac{1}{2}}}\frac{1}{\mu}\frac{1}{L}\sum_{k=0}^{\infty}\frac{16^{-k}}{k!}
\prod_{j=1}^{k}[4-(2j-1)^{2}]\hspace{5cm}\nonumber\\
\times\sum_{n_{1},n_{2},n_{3},n=1}^{\infty}n_{1}^{-k-\frac{3}{2}}
(\pi^{2}n_{2}^{2}+\pi^{2}n_{3}^{2}+\mu^{2}n^{2})^{-\frac{2k-1}{4}}\hspace{2.5cm}\nonumber\\
\times\exp[-2n_{1}(\pi^{2}n_{2}^{2}+\pi^{2}n_{3}^{2}+\mu^{2}n^{2})^{\frac{1}{2}}]\hspace{3cm}\nonumber\\
-\frac{\pi}{96}\frac{1}{\pi}\frac{1}{L}
+\frac{3}{16\pi^{\frac{3}{2}}}\frac{1}{\mu}\Gamma(\frac{3}{2})\zeta(3)\frac{1}{L}\hspace{6.2cm}\nonumber\\
+\frac{\pi^{\frac{1}{2}}}{2}\frac{1}{\mu}\frac{1}{L}\sum_{n_{2},n_{3}=1}^{\infty}\frac{n_{3}}{n_{2}}K_{1}(2\pi
n_{2}n_{3})-\frac{1}{4\pi^{2}}\frac{1}{\mu^{2}}\Gamma(2)\zeta(2)\beta(2)\frac{1}{L}\hspace{1.8cm}\nonumber\\
-\frac{1}{2\mu}\frac{1}{L}\sum_{k=0}^{\infty}\frac{(16\pi)^{-k}}{k!}
\prod_{j=1}^{k}[4-(2j-1)^{2}]\hspace{4.8cm}\nonumber\\
\times\sum_{n_{1},n_{2},n_{3}=1}^{\infty}n_{1}^{-k-\frac{3}{2}}(n_{2}^{2}+n_{3}^{2})^{-\frac{2k-1}{4}}
\exp[-2\pi n_{1}(n_{2}^{2}+n_{3}^{2})^{\frac{1}{2}}]\hspace{0.8cm}
\end{eqnarray}

\noindent The Casimir energy density in unit of $\frac{1}{L}$ for
an equal-edged cube in the presence of one UXD is depicted in
Figure 2. The sign of the energy density keeps negative, in
accordance with the results in [14, 19]. The derivation and
numerical calculation according to (16) show that there also
exists a special ratio $\mu_{f}=1.2$. We find that the Casimir
force is attractive if $\frac{R}{L}<\mu_{f}$ and it is repulsive
at $\frac{R}{L}>\mu_{f}$. Therefore, the Casimir force may be
repulsive for the rectangular cavity with three equal edges in the
Universe with one UXD, in contrast with the same problem in the
four-dimensional Minkowski spacetime. The nature of Casimir force
depends on the compactified universal extra dimension.

The Casimir effect for a cavity in the presence of compactified
universal extra dimensions is different to that in the
four-dimensional spacetime. This means that we give out a new
method for exploring the existence of extra dimensions of our
Universe. Here we have discussed the Casimir effect for a
rectangular cavity with two or three equal edges in the world with
one extra dimension. We derive and calculate the Casimir energy.
We also discuss the nature of Casimir force between boundaries. We
show that there exist influence of extra dimensions on the Casimir
effect. For a $p=2$ equal-edge cavity, the sign of Casimir energy
will become negative if the edges are sufficiently large, meaning
that the edges are larger enough than the radius of UXD or the
energy will be positive. The boundaries repulse each other in the
case of negative Casimir energy. For the $p=3$ equal-edge cavity,
the Casimir energy keeps negative. The Casimir force between
boundaries is attractive if we choose $\frac{R}{L}<1.2$ and it is
repulsive at $\frac{R}{L}>1.2$, which is different from the
results in [14, 19]. The influence from UXD on the Casimir effect
is manifest.

This work is supported by the Basic Theory Research Fund of East
China University of Science and Technology, grant No. YK0127312
and partly supported by the Shanghai Municipal Science and
Technology Commission No.04dz05905.

\newpage
\begin{figure}
\setlength{\belowcaptionskip}{10pt} \centering

  \caption{The curve of Casimir energy density as function of $R/L$
   for $p=2$ equal-edge rectangular cavity in the spacetime with one UXD.}
\end{figure}

\newpage
\begin{figure}
\setlength{\belowcaptionskip}{10pt} \centering

  \caption{The curve of Casimir energy density as function of $R/L$
   for $p=3$ equal-edge rectangular cavity in the spacetime with one UXD.}
\end{figure}


\newpage
\begin{thebibliography}{99}
\bibitem {Kaluza} Kaluza T 1921 \emph{Sitz. Preuss. Akad. Wiss. Phys.
Math.} \textbf{K1} 966
\bibitem {Klein} Klein O 1926 \emph{Z. Phys.} \textbf{37} 895
\bibitem {Green} Green M B, Schwarz J H, Witten E 1987
\emph{Superstring Theory} (Cambridge Univ. Press)
\bibitem {Horova} Horova P, Witten E 1996 \emph{Nucl. Phys.} \textbf{B460}
506
\bibitem {Arkani-Hamed} Arkani-Hamed N, Dimopoulos S, Dvali G 1998
\emph{Phys. Lett.} \textbf{B429} 263
\bibitem {Antoniadis} Antoniadis I 1990 \emph{Phys. Lett.} \textbf{B246} 317\\
Antoniadis I, Arkani-Hamed N, Dimopoulos S, Dvali G 1998
\emph{Phys. Lett.} \textbf{B436} 257
\bibitem {Arkani-Hamed} Arkani-Hamed N, Dimopoulos S, Dvali G 1999
\emph{Phys. Rev.} \textbf{D59} 086004
\bibitem {Randall} Randall L, Sundrum R 1999 \emph{Phys. Rev.
Lett.} \textbf{83} 4690
\bibitem {Randall} Randall L, Sundrum R 1999 \emph{Phys. Rev.
Lett.} \textbf{83} 3370
\bibitem {Casimir} Casimir H B G 1948 \emph{Proc. Nederl. Akad.
Wetenschap} \textbf{51} 793
\bibitem {Plunien} Plunien G, Muller B, Greiner W 1986 \emph{Phys.
Rep.} \textbf{134} 87\\
Elizalde E, Odintsov S D, Romeo A, Bytsenko A A, Zerbini S 1994
\emph{Zeta Regularization Techniques with Applications} (World
Scientific
Publishing Co. Pte. Ltd.)\\
Elizalde E 1995 \emph{Ten Physical Applications of Spectral Zeta
Fnctions} (Springer-Verlag)
\bibitem {Bordag} Bordag M, Mohideen, Mostepanenko V M 2001 \emph{Phys.
Rep.} \textbf{353} 1\\
Milton K A 2001 \emph{The Casimir Effect, Physical manifestations
of zero-point energy} (World Scientific Publishing Co. Pte.
Ltd.)\\
Bordag M, Elizalde E, Kirsten K 1996 \emph{J. Math. Phys.}
\textbf{37} 895\\
Bordag M, Elizalde E, Kirsten K, Leseduarte 1997 \emph{Phys. Rev.}
\textbf{D56} 4896\\
Elizalde E, Bordag M, Kirsten K 1998 \emph{J. Phys.} \textbf{A31}
1743
\bibitem {Maclay}Maclay J G 2000 \emph{Phys. Rev.} \textbf{A61}
052110
\bibitem {Li} Li X, Cheng Hongbo, Li J, Zhai X 1997 \emph{Phys.
Rev.} \textbf{D56} 2155
\bibitem {Cheng} Cheng Hongbo, Li X 2001 \emph{Chin. Phys. Lett.}
\textbf{18} 1163
\bibitem {Cheng} Cheng Hongbo 2002 \emph{J. Phys. A: Math. Gen.} \textbf{35}
2205\\
Elizalde E, Nojiri S, Odintsov S D, Ogushi S 2003 \emph{Phys.
Rev.} \textbf{D67} 063515\\
Cognola G, Elizalde E, Nojiri S, Odintsov S D, Zerbini S 2004
\emph{Mod. Phys. Lett.} \textbf{A19} 1435
\bibitem {Appelquist} Appelquist T, Cheng H C, Dobrescu B A 2001
\emph{Phys. Rev.} \textbf{D64} 035002
\bibitem {Poppenhaeger} Poppenhaeger K, Hossenfelder S, Hofmann S, Bleicher M 2004 \emph{Phys. Lett.} \textbf{B582} 1\\
Saharian A A, Setare M R 2003 \emph{Phys. Lett.} \textbf{B552} 119
\bibitem {Caruso} Caruso F, Neto N P, Svaiter B F,
Svaiter N F 1991 \emph{Phys. Rev.} \textbf{D43} 1300
\end{thebibliography}
\end{document}